\documentclass[11pt]{article}

\usepackage[final]{acl}
\usepackage{multirow}

\usepackage{siunitx}

\usepackage{tikz}
\usetikzlibrary{positioning}

\usepackage{placeins}

\usepackage{times}
\usepackage{latexsym}

\usepackage[T1]{fontenc}

\usepackage[utf8]{inputenc}

\usepackage{microtype}

\usepackage{inconsolata}

\usepackage{graphicx}

\usepackage{booktabs}

\usepackage[normalem]{ulem}

\title{UCRBench: Benchmarking LLMs on Use Case Recovery}

\author{
Shuyuan Xiao$^{*1}$, Yiran Zhang$^{*2}$, Weisong Sun$^{\dagger 2}$, Xiaohong Chen$^{\dagger 1}$, Yang Liu$^2$, Zhi Jin$^3$\\
$^1$East China Normal University, Shanghai, China \\
$^2$Nanyang Technological University, Singapore \\
$^3$Wuhan University, Wuhan, China \\
}

\begin{document}
\maketitle

\begingroup
\renewcommand\thefootnote{}
\footnotetext{* Shuyuan Xiao and Yiran Zhang contributed equally to this work.}
\footnotetext{$\dagger$ Weisong Sun and Xiaohong Chen are the corresponding authors.}
\endgroup

\begin{abstract}
Use cases are widely employed to specify functional requirements, yet existing benchmarks are scarce and face the risk of being misaligned with actual system behavior, similarly limiting the rigorous evaluation of large language models (LLMs) in generating use cases from source code. We address this gap by introducing code-aligned use case benchmarks, constructed through manual validation of both user-goal and subfunction use cases across nine real-world software projects. Using this benchmark, we conduct the first systematic study of LLMs and propose a hierarchical evaluation protocol that assesses actor correctness, name accuracy, path fidelity, and behavioral coverage. The results show that while LLMs can partially reconstruct system functionality, their performance varies significantly across projects, with particularly noticeable shortcomings in domain-specific and multi-module systems. The models also exhibit high omission rates and struggle to maintain consistent abstraction when aggregating subfunctions into user-goal use cases, highlighting both the potential and current limitations of LLM-based use case reverse engineering.
\end{abstract}

\section{Introduction}

Use cases have been widely accepted and acknowledged as a specification method for describing the functional requirements of a software system \cite{TIWARI2015128}. They provide a structured description of interactions between users and systems, enabling developers and analysts to better understand functional requirements. Therefore, high-quality use case benchmarks are crucial for evaluating automated requirement analysis methods and supporting empirical software engineering research.

Existing use case benchmarks are extremely scarce, they can be categorized into use case model benchmarks and textual use case description benchmarks. Model-based benchmarks usually originate from student modeling projects or public UML repositories \cite{remodd2016}. Textual benchmarks are generally produced by manually extracting scenario-like descriptions from narrative artifacts, such as those found in the CMU Scenario Corpus \cite{lee2006scenario}, or by transforming natural-language requirements into simplified use case flows, as in the NL2UseCase dataset \cite{bose2012nl2usecase} and iTrust use case descriptions \cite{itrust2007}. However, these benchmarks are typically loosely coupled with the actual system implementation because they originate from high-level requirement documents, educational examples, or manually constructed scenarios, the underlying textual descriptions often suffer from inconsistencies with the real software behavior. Therefore, a reliable use case benchmark that is explicitly aligned with the source code is essential for ensuring completeness, verifiability, and faithful reflection of actual system behavior.

Recent advances in LLMs have demonstrated impressive capabilities in code comprehension \cite{ding2024semcoder}, cross-artifact reasoning, and requirement-level text generation \cite{beg2025formal}. However, systematic evaluation of LLMs for use case reverse engineering remains largely unexplored. Existing studies that attempt to derive requirements from source code are extremely limited and typically focus on generating high-level user stories from small code fragments rather than constructing structured use cases. For example, a recent effort \cite{ouf2025reverse} investigates producing user stories from isolated C++ methods, but does not address the challenges posed by real-world systems involving multiple modules, interacting components, and non-trivial control flow. Consequently, it remains unclear whether current LLMs can reason over multi-file and multi-module contextual dependencies, maintain consistency across actors and domain entities and generate hierarchical behavioral specifications at different granularity levels. This gap highlights the need for a high-quality, code-aligned benchmark to rigorously assess LLMs’ ability to reverse use cases from full-scale software systems.

In this paper, we construct a manually validated benchmark by annotating use cases from nine real-world software projects of different sizes and domains. We follow established use case theory \cite{cockburn2001} to construct two complementary benchmarks: a user-goal use case dataset, representing high-level functional objectives from the user perspective and a subfunction use case dataset, capturing fine-grained functional units decomposed from the underlying implementation. Each use case includes three essential elements: actor, name and path. For user-goal use cases, we directly extract candidate descriptions from project documentation and then refine and validate them with reference to the corresponding source code. Based on the core source code behind the user-goal use cases, we break the code down into individual subfunctions and derive the subfunction use cases \cite{cockburn2001}.

To evaluate the reverse-engineering capability of LLMs, we choose four representative models (GPT-5, GPT-5 mini, DeepSeek-v3.2 in both non-reasoning mode and reasoning mode) and provide each model with the source code of the projects to generate two-level use cases. The models are first prompted to generate subfunction use cases, and then instructed to aggregate them, based on their understanding of system functionality, into user-goal use cases. This procedure yields two levels of LLM-generated datasets. We then compare LLMs' outputs with the UCRBench by matching them in the order of path → name → actor, and compute similarity scores at both levels. In addition, we measure how many manually curated use cases are missing in the results, providing further insight into coverage and completeness.

Our experimental results reveal substantial variability in the quality of LLM-generated use cases. While the models demonstrate the ability to reconstruct certain functional behaviors, their performance is highly sensitive to project scale, domain complexity, and required granularity. Well-structured projects are interpreted more reliably, whereas specific domain systems and multi-module architectures pose significant challenges. The models also struggle to infer fine-grained actors and often fail to maintain consistent abstraction levels when aggregating subfunction into user-goal use cases. 
Overall, the findings highlight both the potential and the current limitations of LLMs in reversing use cases from source code, underscoring the need for code-aligned benchmarks to guide future progress in automated requirement extraction.

\begin{figure}[t]
    \centering
    \includegraphics[width=\linewidth]{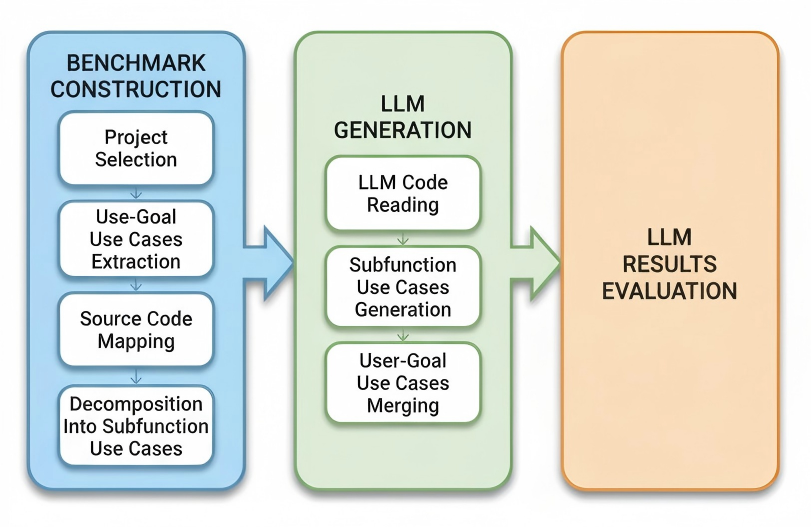}
    \caption{Workflow for Construction, Generation, and Evaluation.}
    \label{fig:Workflow}
\end{figure}

\section{Background}

\subsection{Use Case}
Use case serves as a structured representation of a distinct functional capability of the system. To ensure clarity and consistency across projects, we define every use case using three basic elements: an actor, a name, and an execution path, denoted as the tuple $\langle A, N, P \rangle$, as illustrated in Figure~\ref{fig:format}. The actor and name form the fundamental semantic backbone of the use case, identifying \textit{who} interacts with the system and \textit{what} functionality is invoked \cite{cockburn2001}. The execution path further establishes the mapping between use case and source code.

Given the inherent variability in use case granularity, where the goals and interactions can be unfolded into finer-grained functional elements, we employ a hierarchical constructing approach. Specifically, we define two levels of use cases: user-goal and subfunction use cases \cite{cockburn2001} to accommodate requirements at multiple levels of abstraction.

\subsubsection{User-goal Use Case}
The user goal is the goal of greatest interest. User-goal use cases represent high-level, meaningful objectives that an actor seeks to accomplish through a complete and coherent interaction with the system. It reflects what the user fundamentally comes to the system to achieve, rather than how these objectives are implemented internally. This level provides an essential abstraction layer for understanding system functionality from the perspective of user intent, forming the backbone of requirement analysis and system design.

\subsubsection{Subfuction Use Case}
Subfunction use cases describe lower-level operational tasks or supporting interactions required to fulfill a user-goal use case. These goals are not meaningful standalone achievements for the user, rather, they serve as intermediate steps, reusable components, or auxiliary operations necessary for realizing higher-level objectives. Subfunction-level goals can improve clarity, modularize basic interactions and reflect functional structures shared across multiple user-goal scenarios.



\section{Benchmark Construction}

We follow a multi-stage annotation workflow that integrates documentation analysis, source-code inspection, and exception handling to construct our benchmark. Firstly, we select proper projects to extract and validate user-goal use cases by combining project documentation with their corresponding code regions. After that, we identify the major implementation segments and further decompose them into subfunction use cases, establishing explicit code to use case mappings at both levels.

\subsection{Project Selection}
Java is widely regarded for its versatility and practical utility \cite{redmonk2025ranking, tiobe2025index}. Its prevalence makes it well-suited for studying reverse-engineered use case. Therefore, We selected candidate projects from GitHub by focusing on repositories implemented in Java and ensuring diversity in project scale and domain. To achieve this, we ranked GitHub repositories by their star counts and inspected them sequentially to collect projects covering diverse application domains, ensuring that the benchmark includes systems with different functional focuses. Through this process, we reviewed 200 Java-based repositories and ultimately selected nine representative projects. Several criteria guided our selection process.

First, each project must include a clear and complete README.md file, enabling us to understand the system’s purpose, functionality, and usage before conducting deeper analysis. Second, the project must provide functional documentation or UML use case diagrams, as such high-level design artifacts directly affect the reusability and verifiability of the study by enabling consistent interpretation and reliable extraction of user-goal use cases \cite{GonzalezBarahona2012Reproducibility}. Third, the project organizes the repository with a modular structure, such as grouping components by core functionalities or user roles to enable efficient code navigation and facilitate traceability. These criteria collectively ensure that the selected projects support accurate, consistent, and reproducible construction of both user-goal and subfunction use case datasets.

In total, nine projects were selected to ensure the generality of our benchmark. With the project set finalized, we proceeded to extract user-goal and subfuction use cases from their documentation and source code.

\subsection{Benchmark Construction Procedure}

\begin{figure*}[t]
    \centering
    \includegraphics[width=\linewidth]{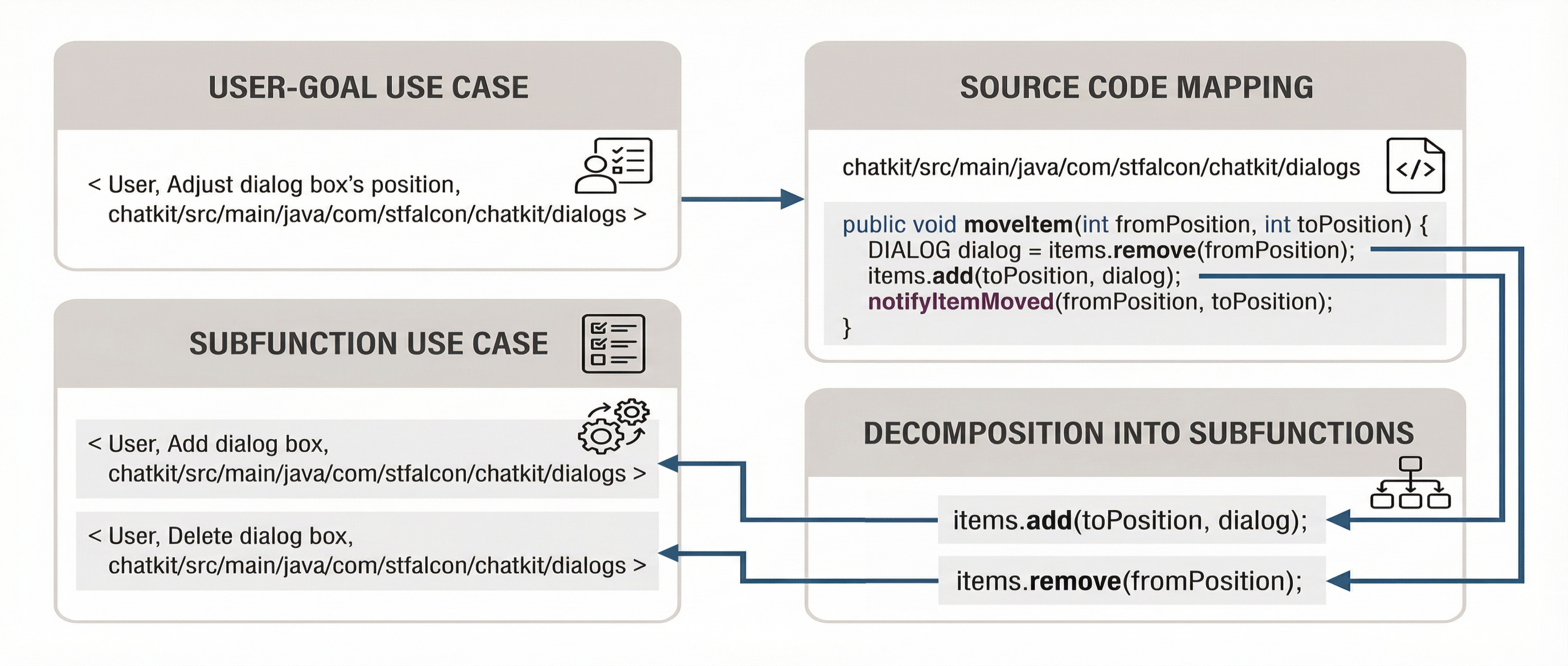}
    \caption{Decomposition of User-Goal Use Cases into Subfunction Use Cases.}
    \label{fig:format}
\end{figure*}

\subsubsection{Use Case Format}
We establish rules for identifying actors, formulating concise yet meaningful use case names, and extracting execution paths that accurately reflect the underlying code logic.

\noindent\textbf{Principle 1 (Actor):} The primary entity, typically including regular users (e.g., \textit{Student}), administrators, and developers, that initiates or participates in the use case. To maintain clarity, each use case is assigned a single primary actor. 

\noindent\textbf{Principle 2 (Name):} The name conveys the core functionality of a use case. We adopt a unified naming convention of “verb + noun” (e.g., \textit{Create Account}) to maintain consistency and readability.

\noindent\textbf{Principle 3 (Path):} Specify the directory location of the .java files that implement the corresponding use case. It serves as a structural linkage between functional descriptions and their concrete code implementations, enabling accurate code to use case mapping.

\subsubsection{User-goal}
The construction of the user-goal benchmark follows a structured procedure that integrates documentation analysis with source code validation.

We begin by reading the project README files, functional descriptions, and any available UML use case diagrams to extract user-level goals, as illustrated in Figure~\ref{fig:format}. These artifacts typically provide high-level descriptions of what the system offers, making them suitable for identifying initial user intentions and main functional categories.

Next, each candidate user goal is examined and refined through source code inspection. By explicitly locating the directory that contains the corresponding implementation files, we establish a concrete mapping between the use case and the source code to support the generated use case \cite{WANG2024102118}. During this stage, we identify all folder that covers .java files required for that use case, establishing a stable path To use case mapping.

Since inconsistencies and omissions in project documentation are common \cite{Mucha2024PreRSTraceability}, it is necessary to verify whether the functionalities described in the documentation are actually implemented in the project. We apply exception handling mechanisms, including removing use case goals that are not supported by the code, correcting the actual actor and name of each use case, and summarizing use cases from implementation files that are not covered by the documentation, thereby aligning the use cases with the real system behavior and ensure the validity and consistency of the benchmark dataset.

After resolving all inconsistencies and completing the necessary refinements, the finalized set of user goals forms the upper layer of our benchmark. Each user-goal use case contains a clearly defined actor, name, and path, ensuring traceability and reproducibility across projects.

\subsubsection{Subfunctions}
Then we construct the subfunction benchmark by decomposing each user goal into finer-grained functional units\cite{cockburn2001} based on its corresponding implementation code, as illustrated in Figure~\ref{fig:format}.

For every user-goal use case, we first locate the core source code regions that implement its main behavior. Once the relevant implementation area is identified, we decompose the user-goal functionality into subfunction goals. Each subfunction corresponds to a coherent operational step, intermediate functional responsibility, or internally reusable behavior reflected in the code. The decomposition process adheres to one principle: each subfunction must represent a meaningful functional unit rather than a low-level technical action.

For each subfunction goal, we annotate the actor, name, and path, establishing a clear mapping between subfunctional steps and source code structure. The final subfunction benchmark provides a detailed representation of system behavior that complements the user-goal abstraction and supports fine-grained evaluation of LLM-based reverse use case generation.

\subsection{Construction Result}
The construction results of our benchmark are presented in Table~\ref{tab:benchmark-overview}, covering both user-goal and subfunction use cases across nine real-world Java projects. The benchmark comprises a diverse set of systems spanning multiple domains, including library management, chat UI tool, AI assistant and so on.
Across these projects, we annotated 232 user-goal use cases and 324 subfunction use cases, with LOC ranging from under 1,000 lines of code to over 60,000 lines. This diversity in domain, scale, and structural complexity ensures that the benchmark offers comprehensive and representative scenarios for evaluating LLMs on reversing use case generation.

\begin{table*}[t]
    \centering
    \small
    \tabcolsep=8pt
    \caption{Overview of UCRBench Benchmarks.}
    \label{tab:benchmark-overview}

    \begin{tabular}{l c c l l}
        \toprule
        \textbf{PROJECT} & \textbf{USER-GOAL} & \textbf{SUBFUNCTION} & \textbf{LOC} & \textbf{DOMAIN} \\
        \midrule
        Library   & 12 & 15 & 863   & Library Management \\
        Chatkit   & 20 & 20 & 4696  & Chat UI Tool \\
        Baseadmin & 26 & 42 & 5488  & Admin Management \\
        Poli      & 15 & 22 & 6200  & Business Intelligence and Reporting \\
        Petclinic & 38 & 58 & 9894  & Pet Clinic Management \\
        Didicar   & 25 & 30 & 10025 & Ride-Hailing \\
        Ruoyi     & 22 & 47 & 12091 & AI Assistant \\
        JetUML    & 45 & 53 & 32452 & UML Modeling Tool \\
        Xpipe     & 29 & 37 & 66609 & Remote Infrastructure Management \\
        \bottomrule
    \end{tabular}
\end{table*}

\section{Evaluation}
Based on our benchmark, we further assess LLMs' capability in reversing two-level use cases. We focus on the following RQs.
\begin{itemize}
    \item \textbf{RQ1:} To what extent can LLMs effectively generate subfunction use cases?

    \item \textbf{RQ2:} To what extent can LLMs effectively generate user-goal use cases?
\end{itemize}

We adopt a two-level evaluation framework. At the lower level, models generate subfunction use cases directly from source code structures; at the higher level, the models merge these subfunction units into coherent user-goal use cases.Then, we design a multi-stage matching and scoring procedure grounded in structural, lexical, functional similarity, and omission count to evaluate the generation quality of LLMs.

\subsection{LLM-based Generation Methods}

\subsubsection{Subfunction}
At the subfunction level, the system extracts fine-grained yet user-relevant functional operations directly from the source code.

The process begins with code fragmentation based on folder, where the system traverses all directories containing Java files while excluding non-functional folders such as \textit{test}, \textit{sample}, and \textit{resource}. If the amount of code within a directory does not exceed the maximum input limit, the directory is treated as an independent functional unit, and all \texttt{.java} files within that directory and its subdirectories are merged and submitted to the LLM as a single analysis chunk. If the code size exceeds the limit, the system recursively applies the same check to the next-level subdirectories and uploads the code in batches. This design prevents exceeding the LLM's input capacity and ensures that structurally related code is analyzed together as a cohesive module.

The LLM then interprets each code chunk and identifies subfunction use cases, each representing a reusable operation that contributes to user-goal workflows. Actors and names are assigned following the same standards used in the human-annotated dataset. The path returned by the model serves as a structural anchor, linking each subfunction to the implementation modules from which it is derived.

After all code units are processed, the system parses the LLM outputs into structured triples $\langle A, N, P \rangle$, which serve as the foundational input for subsequent merging at the user-goal level.

\subsubsection{User-goal}
At the user-goal level, the system aggregates multiple subfunction use cases into higher-level user-goal use cases, each representing a semantically complete and meaningful user operation. To improve the accuracy of merging, we first group subfunction use cases by their actor, ensuring that only use cases initiated by the same participant are considered together. The LLM then analyzes the names of all subfunctions within each group to identify which ones can be combined to jointly achieve a higher-level objective. Only subfunctions that belong to the same operational workflow and involve the same domain entity are eligible for merging; cross-domain or semantically unrelated subfunctions are strictly prohibited. For subfunctions that are successfully merged, the LLM generates a new user-goal name while preserving the original actor.

The system then reconstructs the path for each user goal by aggregating the directory paths of all contributing subfunctions. If a subfunction already represents a complete operational goal on its own, it is directly promoted to the user-goal level without modification. The final outputs, including the generated name, actor, and the list of contributing subfunctions, are exported as structured user-goal benchmark files, ensuring consistency and reproducibility for downstream evaluation.

\subsection{Experimental Setup}
\subsubsection{Model Selection}
We evaluate four representative LLMs for the task of reverse-engineering use cases. Our selection includes GPT-5 and GPT-5-mini, along with DeepSeek-V3.2 in both its non-reasoning mode and reasoning mode. These models differ in scale and inference behavior \cite{NEURIPS2022_9d560961}, allowing us to analyze how model capacity and reasoning capabilities affect their performance in generating subfunction and user-goal use cases.

\subsubsection{Evaluation Metrics}
\paragraph{ Actor Accuracy ($Acc_A$)}
This score measures the accuracy of the actor elements generated by the LLM within a use case. Following our implementation, $Acc_A$ is computed using a hybrid similarity function that combines (1) semantic similarity based on SBERT embeddings \cite{reimers-gurevych-2019-sentence} and (2) category similarity derived from a predefined actor-role taxonomy. This design enables the metric to capture both fine-grained lexical semantics and higher-level role equivalence.

\begin{equation}
    Acc_A = w_{\text{s}} \cdot Sim_{\text{s}}(a',a)
          + w_{\text{c}} \cdot Sim_{\text{c}}(a',a)
\end{equation}

where $a'$ denotes the predicted actor, $a$ is actor from UCRBench, $sim(\cdot)$ denotes the similarity function, $s$ denotes the SBERT embedding function, and $c$ maps an actor token to its role category
(\textit{end\_user}, \textit{privileged\_user}, \textit{system\_staff}). We set $w_{\text{s}} = 0.3$ and $w_{\text{c}} = 0.7$.

\paragraph{Name Accuracy($Acc_N$)}
It assesses whether the model accurately conveys the intended functional goal of a use case through its generated name. Use case names are decomposed into verb phrases and noun phrases, and semantic similarity is computed independently for the verb and noun components based on cosine similarity between normalized sentence embeddings \cite{reimers-gurevych-2019-sentence}. This formulation allows the metric to simultaneously capture alignment at both the action level and the object level of use case descriptions.

\begin{equation}
    Acc_N = w_{\text{v}} \cdot Sim_{\text{n}}(v',v)
          + w_{\text{n}} \cdot Sim_{\text{n}}(n',n)
\end{equation}

where $v'$ and $n'$ represent the verb and noun phrases extracted from the predicted use case name, $v$ and $n$ are those name from UCRBench. We set $w_{\text{v}} = 0.5$ and $w_{\text{n}} = 0.5$ based on preliminary experiments.

\paragraph{Path Accuracy($Acc_P$)}
It evaluates whether LLMs can correctly identify the source-code location. We compare the LLM-produced folder path with the ground-truth path using Jaccard similarity over directory segments.The path score is the similarity value derived from these structural comparisons.
\begin{equation}
    Acc_P = \frac{|p' \cap p|}{|p' \cup p|}
\end{equation}

where $p'$ represents the predicted path combination and $p$ is the path combination from UCRBench.

\paragraph{Omission Rate($OR$)}
This rate measures the proportion of UCRBench that the model fails to match, relative to the total number of annotated use cases. As a complementary metric to accuracy scoring, it reflects LLMs' ability to achieve sufficient coverage over the functional space represented in the benchmark.
\begin{equation}
    OR = \frac{\#\text{Omissions}}{\#\text{Use Cases}}
\end{equation}

where \#\text{Omissions} denotes the number of unmatched use cases and \#\text{Use Cases} indicates the total use cases in UCRBench.

\subsection{Automated Evaluation Procedure}

To systematically evaluate the generation quality of LLMs, we adopt a sequential, multi-stage matching pipeline that operates in a fixed order: path → name → actor comparison, followed by an assessment of omission cases. This top-down procedure ensures that only candidate use cases originating from the most similar source code regions are allowed to undergo deeper lexical and semantic evaluation, and that each actor matching decision is grounded in the similarity of name.

At the matching stage of path, We compute $Acc_P$ between the model-generated path set and UCRBench path set. Only the use cases with the highest $Acc_P$ score are retained as candidates for the next matching stage. This ensures that only use cases originating from nearly identical execution paths proceed to the subsequent semantic comparison, preventing cases where semantically similar descriptions are matched despite representing different underlying functionalities.

For all candidates with the highest path similarity, we further compare the semantic similarity of their use case names. The human use case with the highest $Acc_N$ score is selected as the final match, ensuring that use cases describing similar functionalities are correctly aligned.

After determining the best use case match, we compute the $Acc_A$. This step evaluates whether the model assigns the correct role to each use case and whether it can identify special actors based on the underlying code.

At this stage, each use case is assigned its uniquely matched UCRBench counterpart and receives the corresponding path, name, and actor scores. Use cases in UCRBench that fail to be matched by any LLM-generated use cases are counted and reported as $OR$. These results are presented in Table~\ref{tab:llm-comparison-subfuction} and Table~\ref{tab:llm-comparison-user-goal}.

\subsection{Research Questions}
\subsubsection{RQ1: To what extent can LLMs effectively generate subfunction use cases?}
Table~\ref{tab:llm-comparison-subfuction} summarizes how LLMs perform on subfunction use case reverse engineering, providing a comparative view across different models.

\paragraph{$Acc_A$}
Across all projects, LLMs achieve average $Acc_A$ scores ranging from 62.2 to 77.2 in actor extraction. All models exhibit similar challenges in identifying appropriate actors for low-level system behaviors. In function-oriented projects such as \textit{Baseadmin} and \textit{Didicar}, the models frequently confuse user-level and developer-level actors, resulting in incorrect actor assignments.

GPT-5 achieves higher $Acc_A$ scores on several projects, including \textit{Chatkit} and \textit{Poli}. This improvement is primarily associated with its ability to correctly assign \textit{developer} as the actor for a large portion of implementation-level use cases. However, mismatches still occur for use cases where \textit{user} is the expected actor. On the \textit{Library} project, three models obtain relatively higher scores by correctly identifying the \textit{librarian} role. In contrast, DS-R fails to recognize this actor. Nevertheless, none of the models consistently distinguish more fine-grained actors such as \textit{student} and \textit{teacher}.

\paragraph{$Acc_N$}
All models achieve relatively low scores in use case name extraction. This is mainly because the models tend to generate an excessive number of subfunction-level use cases, many of which correspond to overly low-level operations. Such over-generation leads to a large portion of the generated use cases being unmatched with the reference set during evaluation. For example, in the \textit{Xpipe} project, GPT-5 generates as many as 2103 subfunction use cases, while DS-R produces 887, many of which correspond to fine-grained behaviors such as ``Check GPU'' or ``Show loading.'' These low-level actions are difficult to align with the manually annotated reference use cases, ultimately lowering the overall $Acc_N$ scores of the models.

\paragraph{$Acc_P$}
DeepSeek outperforms GPT on eight projects. DeepSeek tends to extract independent use cases for nearly every execution path, representing subfunctions in a more fine-grained manner. In contrast, GPT begins to aggregate paths into more comprehensive functionalities at this level, often resulting in a larger number of paths than those in the reference set. This behavior explains why DeepSeek achieves full accuracy on \textit{Library}.  

However, DeepSeek obtains noticeably lower performance on \textit{Ruoyi}. This project exhibits a deeper and more layered code structure, which requires stronger capability in associating and composing related execution paths. In such settings, DeepSeek tends to overlook code segments that are relevant but less explicitly expressed, leading to incomplete path reconstruction.

\paragraph{$OR$}
DeepSeek achieves an average $OR$ of 34, while GPT attains 34.5. Although both model families extract a large number of subfunction-level use cases, a substantial portion of them fails to achieve effective matches. GPT retains more subfunction candidates; however, many of these use cases are directly derived from function names, resulting in numerous low-level operations that cannot be regarded as user-oriented functionalities. Due to their excessively fine granularity, not all extracted use cases translate into better coverage of the reference subfunction set. This behavior leads to relatively low $OR$ on \textit{Xpipe}, while limiting overall performance on projects such as \textit{Baseadmin} and \textit{Poli}.

\subsubsection{RQ2: To what extent can LLMs effectively generate user-goal use cases?}
The performance of different LLMs on user-goal use case reverse engineering is presented in Table~\ref{tab:llm-comparison-user-goal}.

\paragraph{$Acc_A$} 
The DeepSeek family achieves slightly higher average scores than the GPT family, with an overall score of 78.4 for DeepSeek compared to 73.8 for GPT. GPT tends to retain a larger number of low-level subfunctions that cannot be effectively merged, leading to granularity mismatches and lower scores, as particularly evident in \textit{Chatkit}. We also observe that, at this level, GPT is able to identify some finer-grained user roles, such as \textit{passenger} and \textit{owner}, whereas DeepSeek more frequently recognizes higher-level privileged users, such as \textit{admin}.

\paragraph{$Acc_N$}
For the use case name metric, the four models exhibit relatively balanced performance, with scores ranging from 35.5 to 43.4 and only minor variation across projects. This pattern mainly stems from the LLMs' limited understanding of user-level business logic, which prevents many subfunction-level use cases from being properly merged and abstracted into high-level use cases. Both overly broad and overly fine-grained summarization lead to lower scores. For example, GPT-5 outperforms the other three models on \textit{Petclinic} because it avoids excessive aggregation under the ``Manage….'' prefix and instead distinguishes a broader set of business objectives. In contrast, DeepSeek achieves higher scores on \textit{Library}, where the smaller codebase makes aggregation under ``Manage….'' more appropriate.

\paragraph{$Acc_P$}
The performance of all four model groups on this score is not high. In contrast, DeepSeek can more appropriately merge subfunction use cases into single user-goal use cases, achieving better alignment with the path combinations present in the human-annotated dataset. GPT-5 may exhibit over-abstraction, resulting in an excessive number of paths in the path combinations, as seen in the \textit{baseadmin} and \textit{jetuml} projects. DeepSeek, when facing hierarchically complex code, may discard some subpaths and interpret entire large files as a whole, leading to results that are neither precise nor comprehensive, as observed in the \textit{rouyi} project.

\paragraph{$OR$}
The omission rate is notably high, with average values ranging from 49\% to 64\%. This indicates that, although LLMs can partially reconstruct actors, names, and paths, a substantial portion of user-goal use cases is still not captured at all, reflecting a major limitation in the current capabilities of LLMs. In the \textit{rouyi} project, the difference is particularly pronounced. The DS-R model achieves an omission rate of 9\%, with a relatively rich combination of code files covered at the user-goal level. In contrast, the other three models tend to over-abstract; notably, GPT-5 even merges 19 subfunction use cases into a single user-goal use case.

\begin{table*}[t]
    \centering
    \footnotesize
    \tabcolsep=3pt
    \caption{LLM performance on subfuction use case reverse engineering. G5 = GPT-5,\;
    G5m = GPT-5mini,\;
    DS-C = DeepSeek-V3.2-Chat,\;
    DS-R = DeepSeek-V3.2-Reasoner.}
    \label{tab:llm-comparison-subfuction}

    \begin{tabular}{lccccccccccccccccc}
    \toprule
    
    & \multicolumn{4}{c}{$Acc_A$}
    & \multicolumn{4}{c}{$Acc_N$}
    & \multicolumn{4}{c}{$Acc_P$}
    & \multicolumn{4}{c}{$OR$(\%)} \\
    
    \cmidrule(lr){2-5}\cmidrule(lr){6-9}\cmidrule(lr){10-13}\cmidrule(lr){14-17}
    
    & {G5} & {G5m} & {DS-C} & {DS-R}
    & {G5} & {G5m} & {DS-C} & {DS-R}
    & {G5} & {G5m} & {DS-C} & {DS-R}
    & {G5} & {G5m} & {DS-C} & {DS-R} \\
    
    \midrule
    
    Library   & 100.0 & 96.8 & 97.1 & 65.6 & 44.6 & 44.7 & 80.0 & 80.8 & 43.3 & 41.8 & 100.0 & 100.0 & 27 & 33 & 20 & 20 \\
    Chatkit   & 92.1  & 74.8 & 67.3 & 73.4 & 40.9 & 39.0 & 55.2 & 50.5 & 66.7 & 91.7 & 98.2 & 100.0 & 50 & 20 & 45 & 50 \\
    Baseadmin & 56.0  & 58.4 & 56.8 & 57.7 & 25.7 & 26.2 & 28.9 & 29.0 & 77.5 & 77.2 & 80.2 & 77.0 & 69 & 48 & 52 & 55 \\
    Poli      & 84.7  & 51.1 & 47.2 & 41.0 & 32.2 & 27.5 & 22.9 & 21.8 & 79.1 & 68.8 & 86.2 & 92.2 & 68 & 36 & 59 & 59 \\
    Petclinic & 82.1  & 79.7 & 91.5 & 96.1 & 63.4 & 53.8 & 56.5 & 58.1 & 54.2 & 61.8 & 77.1 & 77.6 & 45 & 40 & 47 & 47 \\
    Didicar   & 53.5  & 43.8 & 48.0 & 46.6 & 30.7 & 33.4 & 35.0 & 35.0 & 71.8 & 84.0 & 92.0 & 91.8 & 30 & 10 & 23 & 23 \\
    Rouyi     & 79.0  & 66.4 & 29.2 & 30.6 & 35.9 & 35.0 & 15.6 & 15.9 & 42.8 & 60.3 & 33.5 & 33.8 & 55 & 34 & 32 & 19 \\
    Jetuml    & 69.7  & 71.3 & 82.5 & 82.4 & 28.1 & 29.0 & 35.1 & 34.8 & 52.9 & 67.6 & 88.5 & 89.6 & 36 & 11 & 26 & 19 \\
    Xpipe     & 78.1  & 78.2 & 62.3 & 66.1 & 26.6 & 26.8 & 24.2 & 25.0 & 44.8 & 51.9 & 55.5 & 58.7 & 5  & 5  & 8  & 8  \\
    \midrule
    Average   & 77.2  & 69.0 & 64.7 & 62.2 & 36.4 & 35.0 & 39.3 & 39.0 & 59.2 & 67.2 & 79.0 & 80.1 & 43 & 26 & 35 & 33 \\

    \bottomrule
    \end{tabular}

\end{table*}

\begin{table*}[t]
    \centering
    \footnotesize
    \tabcolsep=3pt
    \caption{LLM performance on user-goal use case reverse engineering. G5 = GPT-5,\;
    G5m = GPT-5mini,\;
    DS-C = DeepSeek-V3.2-Chat,\;
    DS-R = DeepSeek-V3.2-Reasoner.}
    \label{tab:llm-comparison-user-goal}
    
    \begin{tabular}{lccccccccccccccccc}
    \toprule
    
    & \multicolumn{4}{c}{$Acc_A$}
    & \multicolumn{4}{c}{$Acc_N$}
    & \multicolumn{4}{c}{$Acc_P$}
    & \multicolumn{4}{c}{$OR$(\%)} \\
    
    \cmidrule(lr){2-5}\cmidrule(lr){6-9}\cmidrule(lr){10-13}\cmidrule(lr){14-17}
    
    & {G5} & {G5m} & {DS-C} & {DS-R}
    & {G5} & {G5m} & {DS-C} & {DS-R}
    & {G5} & {G5m} & {DS-C} & {DS-R}
    & {G5} & {G5m} & {DS-C} & {DS-R} \\
    
    \midrule
    
    Library    & 100.0 & 95.0 & 96.8 & 65.8 & 31.5 & 37.8 & 77.9 & 78.8 & 46.4 & 41.4 & 72.7 & 72.7 & 50 & 42 & 33 & 33 \\
    Chatkit    & 60.4  & 59.2 & 100.0 & 100.0 & 35.9 & 43.4 & 59.0 & 66.4 & 56.9 & 76.7 & 80.0 & 88.0 & 65 & 80 & 85 & 75 \\
    Baseadmin  & 55.5  & 64.6 & 66.2 & 61.8 & 33.3 & 30.4 & 35.0 & 32.4 & 39.9 & 51.5 & 84.6 & 75.1 & 65 & 38 & 54 & 54 \\
    Poli       & 76.0  & 65.1 & 65.8 & 69.6 & 36.2 & 36.4 & 40.9 & 38.8 & 76.4 & 64.0 & 62.0 & 71.7 & 67 & 40 & 20 & 33 \\
    Petclinic  & 83.5  & 78.0 & 94.3 & 88.7 & 65.2 & 47.4 & 52.0 & 47.2 & 55.3 & 52.5 & 53.9 & 58.7 & 61 & 71 & 74 & 63 \\
    Didicar    & 46.4  & 50.8 & 51.3 & 49.1 & 36.8 & 34.8 & 27.7 & 30.5 & 52.1 & 58.7 & 89.2 & 90.9 & 72 & 68 & 28 & 36 \\
    Rouyi      & 59.1  & 72.8 & 86.8 & 81.2 & 27.7 & 27.5 & 28.8 & 26.7 & 36.6 & 24.4 & 25.7 & 24.6 & 73 & 68 & 59 & 9  \\
    Jetuml     & 89.3  & 100.0 & 83.6 & 83.0 & 28.9 & 31.0 & 31.0 & 28.2 & 54.3 & 61.4 & 85.8 & 67.8 & 80 & 82 & 16 & 84 \\
    Xpipe      & 90.7  & 81.5 & 80.9 & 86.4 & 30.2 & 30.4 & 25.9 & 41.2 & 29.7 & 26.0 & 48.7 & 24.9 & 52 & 79 & 31 & 90 \\
    
    \midrule
    Average    & 73.4  & 74.1 & 80.6 & 76.2 & 36.2 & 35.5 & 42.0 & 43.4 & 49.7 & 50.7 & 67.0 & 63.8 & 65 & 63 & 44 & 53 \\

    \bottomrule
    
    \end{tabular}

\end{table*}

\section{Discussion}

Our evaluation reveals substantial variation in the quality and accuracy of LLM-generated use cases. Although the models are capable of reconstructing portions of system behavior, their outputs remain unstable across projects and functional granularity. The issue is particularly pronounced at higher abstraction levels, where contextual reasoning and merging capabilities remain insufficient.

First, we observe that well-structured projects are significantly easier for LLMs to interpret. Projects with clear module boundaries, shallow functional hierarchies, and simple control flows exhibit more direct mappings between code structure and use case semantics. In contrast, large scale systems involve deeply nested logic, cross-module dependencies, and complex workflow interactions, all of which substantially increase the difficulty of recovering accurate use cases from code alone.

LLMs still struggle with inferring actors. They sometimes fail to distinguish coarse-grained categories such as \textit{user} and \textit{developer}, and frequently fail to identify more specific roles embedded in the codebase, such as \textit{admin}, \textit{student}, or \textit{passenger}. The models tend to default to generic actors, suggesting that accurate actor attribution requires a deeper understanding of project-specific behavior patterns and domain conventions.

Domain-specific projects pose significant semantic challenges. Professional domains require awareness of domain regulations, established workflows, and contextual constraints. Without such domain grounding, LLMs often misinterpret functional intent or oversimplify critical logic, leading to incomplete user goal descriptions or incorrect actor assignments. These findings indicate the importance of domain-adaptive fine-tuning or structured domain knowledge infusion when applying LLMs to specialized software systems.

The performance disparity between subfunction and user-goal use cases further highlights the limitations of LLMs in controlling functional granularity. At the subfunction level, models typically rely on function names and local code patterns to infer relatively accurate low-level behaviors. However, they often suffer from overly fine-grained granularity, making it difficult to extract user-oriented use cases. At the user-goal level, models may over-abstract multiple subfunctions into overly broad intentions or fail to merge related operations, resulting in fragmented and inconsistent use cases. Nevertheless, semantic understanding enables better reverse-engineering performance and allows the models to infer more fine-grained actors. This inconsistency indicates that current models lack explicit mechanisms for regulating abstraction levels or reasoning hierarchies. Future work could explore structured prompting or code-guided decomposition methods to achieve more stable granularity control.

While LLMs demonstrate promising potential for automated use case extraction, their performance remains sensitive to project scale, domain complexity, and granularity requirements. Human-constructed datasets continue to play an essential role in identifying model weaknesses and guiding future improvements. Human annotators naturally incorporate domain knowledge, implicit requirements, and project conventions, enabling more coherent interpretations of business logic and execution flows. UCRBench provides valuable supervision signals for enhancing the reasoning capabilities of future LLMs.

\section{Conclusion}

In this paper, we construct a manually validated and code-aligned benchmark for use case reverse engineering, and simultaneously evaluate the ability of LLMs to generate use cases for full-scale software systems. We build two complementary datasets: a user-goal benchmark that captures high-level functional objectives, and a subfunction benchmark that represents low-level implementation behaviors. These benchmarks enable systematic and multi-granular evaluation of model performance. Based on nine real-world projects and four representative LLMs, we conduct a comprehensive analysis of how models reconstruct actors, functional intents, and execution paths from source code.

The experimental results reveal both the potential and the limitations of current models. LLMs are capable of recovering portions of system behavior, particularly in well-structured projects. However, they exhibit notable weaknesses in handling specific domain logic, identifying correct actors, and maintaining consistent abstraction levels when aggregating subfunction into user-goal use cases. These observations highlight fundamental challenges in code understanding, cross-artifact reasoning, and hierarchical requirement reconstruction.

UCRBench provides a rigorous and reproducible foundation for future research on automated requirement extraction. It helps expose concrete weaknesses in existing models and points toward several promising improvement directions, including domain-adaptive training, structured reasoning strategies, and code-guided functional decomposition for better granularity control. We hope that this benchmark will facilitate continued progress at the intersection of software engineering and LLMs.

\bibliographystyle{acl_natbib}
\bibliography{acl}

\appendix

\end{document}